\documentstyle[11pt,epsfig]{article}
\begin{document}
\begin{titlepage}
\vspace*{5mm}
\begin{center} {\large \bf Exact Solution of an Exclusion Model
in the Presence of a Moving Impurity on a Ring} \\
\vskip 1cm
\centerline {\bf Farhad H Jafarpour \footnote
{e-mail: JAFAR@theory.ipm.ac.ir}} \vskip 1cm
{\it  Department of Physics, Sharif University of Technology, }\\
{\it P.O.Box 11365-9161, Tehran, Iran }\\
{\it  Institute for Studies in Theoretical Physics and Mathematics,}\\
{\it P.O.Box 19395-5531, Tehran, Iran}
\end{center}
\begin{abstract}
We study a recently introduced model [8,9] which consists of positive
and negative particles on a ring. The positive
(negative) particles hop clockwise (counter-clockwise) with rate $1$
and oppositely charged particles may swap their positions with
asymmetric rates $q$ and $1$. In this paper we assume that a finite
density of positively charged particles $\rho$ and only one negative
particle (which plays the role of an impurity) exist on the ring.
It turns out that the canonical partition function of this model can
be calculated exactly using Matrix Product Ansatz (MPA) formalism.
In the limit of infinite system size and infinite number of positive particles,
we can also derive exact expressions for the speed of the positive and
negative particles which show a second order phase transition at
$q_c=2\rho$. The density profile of the positive particles on the ring
has a shock structure for $q \leq q_c$ and an exponential behaviour
with correlation length $\xi$ for $q \geq q_c$. It will be shown that
the mean-field results become exact at $q=3$ [8] and no phase transition occurs
for $q > 2$.  \\
\end{abstract}
 \end{titlepage}
\newpage
\section{Introduction}
One-dimensional driven diffusive systems have provided a rich frame-work
for the study of many interesting phenomena such as phase transitions,
shock structures and the spontaneous breaking of translational invariance
in the context of the non-equilibrium statistical mechanics [1-4]. Moreover,
from the Mathematical point of view, some of them are related to the
integrable quantum chain Hamiltonians and can be mapped to the other
non-equilibrium models,
for instance, interface growth and traffic flows [5,6].\\
In this paper, we study a recently introduced exclusion model which
exhibits both phase transition and spatial condensation of particles
[8,9]. In this model two types of particles, called $positive$ and $negative$,
occupy the sites of a periodic one-dimensional lattice of length $L$. The
particles are subjected to hard-core exclusion, so that there are three
possible states at each site: empty, occupied by a positive particle, or
by a negative one. The positive (negative) particles hop to their
immediate right (left) site with rate $1$ provided that it is empty.
Adjacent positive and negative particles also exchange their positions
with asymmetric rates $q$ and $1$. Therefore, during the infinitesimal time
step
$dt$, any bond $(i,i+1)$ $($with $1 \leq i \leq L)$ evolves as follows:
\begin{eqnarray}
(+)(0) & \rightarrow & (0)(+) \ \ \  with \ \ \  rate \ \ \  1  \nonumber \\
(0)(-) & \rightarrow & (-)(0) \ \ \  with \ \ \  rate \ \ \  1   \\
(+)(-) & \rightarrow & (-)(+) \ \ \  with \ \ \  rate \ \ \  q \nonumber \\
(-)(+) & \rightarrow & (+)(-) \ \ \  with \ \ \  rate \ \ \  1  \nonumber
\end{eqnarray}
With periodic boundary conditions, the stationary state of this model has
been studied by Monte-Carlo simulation, mean-field approximation, and
recently
using analytical approaches in the neutral case in which the density of the
positive and negative particles are equal [8-11]. It has been shown
that in the neutral case the model possesses two different phases depending
on the reaction rate $q$ [10], in contrast to the analytical results of [11] which
indicate that there is no such a transition. This model has also been studied
with open boundaries in two special limits [12].\\
Here we study this model in the charged case in which a finite
density of positive particles $\rho$ and only one negative particle are
present on the ring. Using the Matrix Product
Anstaz (MPA) formalism first introduced in [7], we can derive the exact analytic
expression for the partition function of this model. Exact calculations
show that in the limit of infinite system size and infinite number of positive particles
the model has two different phases for $q < 2$, and the transition point
$q_c$ depends on the density of the positive particles. Exact expressions
for the speed of the positive and negative particles can also be
computed in this limit. We will see that these
expressions have different behaviours depending on the value of $q$. For
$q \leq q_c$, the speed of the positive and negative particles, as a
function of $q$, increase linearly with $q$ from zero. However, for $q \geq
q_c$ the speed of the positive particles remains constant while the speed
of the negative particle still increases as a function of both $\rho$ and
$q$. The speed of particles as a function
of $q$ is continuous at the transition point although its derivative
changes discontinuously at this point. Also for $q \geq q_c$ it turns out
that the density profile of the positive
particles has an exponential behaviour with a characteristic length
$\xi=|\ln\frac{q_c}{q}|^{-1}$, while the system presents a shock, i.e. a
sharp discontinuity between a region of high density of particles and a
region of low density for $q \leq q_c$. At the transition point the
correlation length diverges and the model shows
a second order phase transition. We will show that at $q=3$ the
stationary probability of all possible configurations become equal. At this
point no correlation exists and mean-field results are exact [8]. \\
This paper is organized as follows. In the section 2, we obtain the exact
expression for the canonical partition function of the model using the MPA. In
section 3, we derive exact expressions for the speed of both kinds of
particles, and also for the density profile of the positive particles on the ring.
In the last section we compare our results with
those obtained from the neutral case.
\section{Expression of the canonical partition function using the MPA}
According to the MPA formalism [7], the stationary probability distribution
$P(\{ C \})$ of any configuration $\{ C \}$ can be expressed as a trace of a
product of non-commuting operators. For the model proposed here we
assume that there are $M$ positive particles and only one negative
particle on a ring on length $L$. Since this model is translationally
invariant, we can always keep the single negative particle at site $L$
and write the normalized stationary probability distribution as
\begin{equation}
P(\{C\})=\frac{1}{Z_{L,M}}Tr(\prod_{i=1}^{L-1}(\tau_iD+(1-\tau_i)E)A)
\end{equation}
where $\tau_i=1$ if the site $i$ is occupied by a positive particle and
$\tau_i=0$ if it is empty. The matrices $D$, $A$ and $E$ which stand for
the presence of a positive, negative particle, and a hole satisfy the
following algebra introduced in [10]
\begin{eqnarray}
qDA-AD &=& D+A\\
DE &=& E\\
EA &=& E.
\end{eqnarray}
The normalization factor $Z_{L,M}$, which plays the role analogous to the
partition function in equilibrium statistical mechanics, ensures that
$\sum_{all \ conf.}P(\{C\})=1$ and can be written as a trace
\begin{equation}
Z_{L,M}=Tr(G_{L,M}A)=Tr( \sum_{
\{\tau_i=0,1\}} \delta(M- \sum_{i=1}^{L-1} \tau_i)
\prod_{i=1}^{L-1}( \tau_i D+(1- \tau_i)E)A).
\end{equation}
Here $\delta(x)$ is the Kronecker $\delta_{x,0}$. In fact, the expression (2)
is a conditional probability distribution which gives the probability
of finding configuration $\{ C \}$ in the stationary state provided that a
negative particles exists at site $L$. We will see that all the physical
quantities such as the speed of particles and the density profile of the
positive particles can be written in terms of $Z_{L,M}$.
It can easily be checked that a one-dimensional representation of the
algebra (3-5) exists
$$
D=A=E=1
$$
for $q=3$ [8]. In this case (as can be seen from (2)) all
configurations of the system occur with equal probabilities
$P(\{ C \})=\frac{1}{Z_{L,M}}$ and no correlation exists. It implies
that at $q=3$ the mean-field results are exact. We will see that
in this case $Z_{L,M}$ is equal to the total number of all possible configurations of
the system.
Traditionally, one has to find the representations of the algebra
appears in MPA formalism and use it to calculate both the partition
function and physical quantities. But in this paper we show that (see
Appendix) the expression (6) can be calculated directly using the
algebra (3-5) and has a closed form
\begin{equation}
Z_{L,M}=\sum_{i=0}^{M}\frac{(q-3)(\frac{2}{q})^i+1}{q-2}C_{L-i-2}^{M-i}
\end{equation}
in which $C_{i}^{j}=\frac{i!}{j!(i-j)!}$ is the binomial coefficient. At
$q=3$ we have
$$
Z_{L,M}=\sum_{i=0}^{M}C_{L-i-2}^{M-i}=C_{L-1}^{M}
$$
which is nothing except the total number of possible configurations of the
model.
\section{Exact Physical Quantities}
\begin{figure}[!b]
\hspace{4cm}
\psfig{file=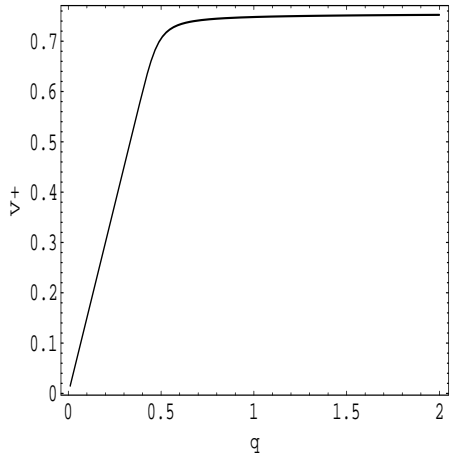,height=6cm,width=6cm}
\caption{ The speed of the positive particles $V_{+}$ as a function
          of $q$ for $\rho=0.25$ (Plot of the Eq.(8)). The
          phase transition takes place at $q_c=0.5$. }
\end{figure}
In the stationary state, the speed of the positive particles in the
reference frame of the lattice is found to be (see Appendix)
\begin{equation}
V_{+}=\frac{(L-M)Z_{L-1,M-1}+C_{L-2}^{M-1}}{MZ_{L,M}}.
\end{equation}
Similarly, the speed of the negative particle in the reference frame of
the lattice can be obtained (see Appendix)
\begin{equation}
V_{-}=\frac{Z_{L-1,M-1}+C_{L-2}^{M}+C_{L-2}^{M-1}}{Z_{L,M}}.
\end{equation}
\begin{figure}[!t]
\hspace{4cm} \psfig{file=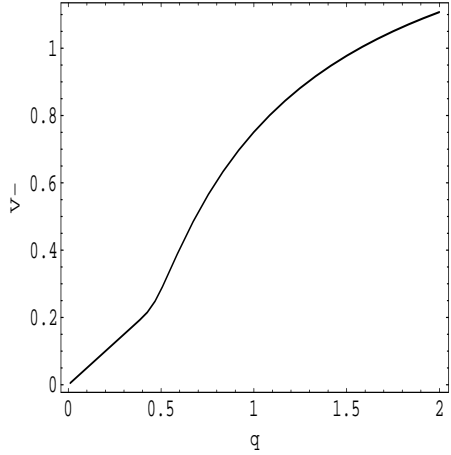,height=6cm,width=6cm} \caption{
The speed of the negative particle $V_{-}$ as a function
          of $q$ for $\rho=0.25$ (Plot of the Eq.(9)). The
          phase transition takes place at $q_c=0.5$.}
\end{figure}
In figures 1 and 2 we have plotted $V_{+}$ and $V_{-}$, respectively,
computed from the exact expressions (8) and (9) as a function of $q$ for
$L=500$ and $M=125$. As can be seen the behaviour of these functions
changes near $q_c=2\rho$. However, we should take the limit $(L,M
\rightarrow \infty$ and $\frac{M}{L-1}=\rho$ being fixed) to find the
exact transition point. We use the steepest descent method for computing
the transition point. At this limit the asymptotic behaviours of (8) and
(9) are given by
\begin{equation}
V_{+}= \left\{
\begin{array}{ll}
\frac{q}{2}(\frac{1-\rho}{\rho})  & \mbox{ if }2\rho \geq q \\
1-\rho                            & \mbox{ if }  2\rho \leq q
\end{array}
\right. ,
\end{equation}
\begin{equation}
V_{-}= \left\{
\begin{array}{ll}
\frac{q}{2}  & \mbox{ if }  2\rho \geq q \\
\rho+\frac{1}{1+\frac{\rho}{q-2}-\frac{\rho(q-2)}{q-2\rho}}  & \mbox{
if }  2\rho \leq q. \end{array}
\right.
\end{equation}
Thus, the transition point is exactly found to be $q_c=2\rho$. Now it is
obvious that for $q > 2$ no phase transition can take place because of
the restriction on the density $\rho < 1$. \\
To study the nature of these phases we can compute the density profile
of the positive particles on the ring. The average density of the positive
particles $n(x)$ at the distance $x$ from the negative particle is equal
to the probability of finding a positive particle at site $L-x-1$.
Using the same procedure proposed in the Appendix, it can be shown that
\begin{equation}
n(x)=\frac{Z_{L-1,M-1}-\frac{q-3}{2}\{ \sum_{i=1}^{M-x} (\frac{q}{2})^{x+i}
C_{L-(x+i)-2}^{M-(x+i)} \}\Theta(M \geq x)}{Z_{L,M}}
\end{equation}
where
\begin{equation}
\Theta(y \geq x )= \left\{
\begin{array}{ll}
1   & \mbox{ if }  y \geq x \\
0   & \mbox{    }  otherwise.
\end{array}
\right.
\end{equation}
As can be seen from (12) the density profile of positively charged
particles is completely flat at $q=3$. In what follows we
study the density profile (12) in the thermodynamic limit.
\begin{figure}[!t]
\hspace{4cm}
\psfig{file=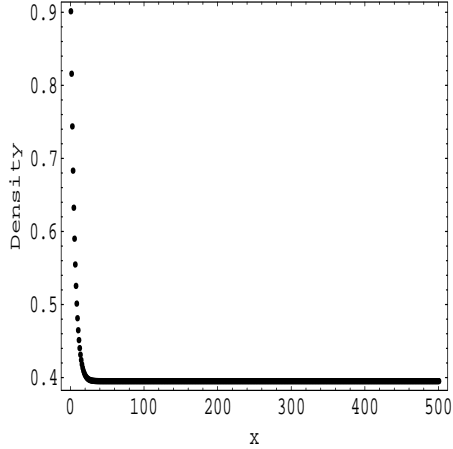,height=6cm,width=6cm}
\caption{The density profile of the positive particles $n(x)$ for $L=500$,
         $M=200$ and $q=0.95$ (Plot of the Eq.(12)). The negative particle is at $x=0$.}
\end{figure}
\subsection{Power-law phase; $2\rho \leq q$}
In this phase, as can be seen from figure 3, the negative particle has a
short-range effect on the system. The density profile starts from $n(0)$
just in front of the negative particle and decreases exponentially to
its bulk value $\rho$. The asymptotic behaviour of formula (12) in the
limit $(L,M \rightarrow \infty$ and $\frac{M}{L-1}=\rho$ being fixed) is given by
\begin{equation}
n(x)=\rho(1-\frac{(1-\rho)(q-2)(q-3)}{(q-2)(q-2\rho)-
\rho(q-2\rho)-\rho(q-2)^2
}e^{-\frac{x}{\xi}})
\end{equation}
in which $\xi=|\ln\frac{q_c}{q}|^{-1}$ specifies the disturbance due to
the existence of the negative particle.
\subsection{Jammed phase; $2\rho \geq q$}
In this domain, the negative particle provokes a macroscopic shock in the
system, as can be seen in the figure 4. A high density region $\rho_{High}=1$
extending from $x=0$ to $x=x_0$ is separated by a sharp interface from a
region of low density $\rho_{Low}=\frac{q}{2}$ extending from $x=x_0$ to
$x=L-1$. The position of the shock $x_0$ can be obtained from the
thermodynamic behaviour of (12) in this region
$$x_0=L\frac{q-2\rho}{q-2}.$$
\begin{figure}[!t]
\hspace{4cm}
\psfig{file=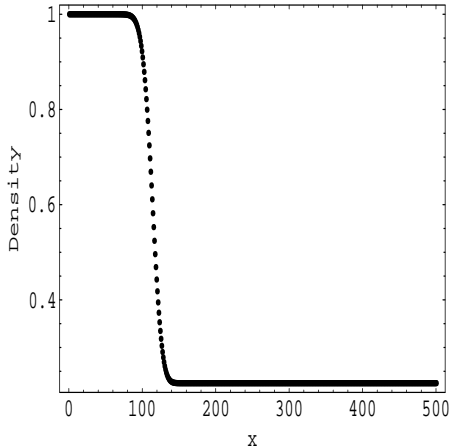,height=6cm,width=6cm}
\caption{ The density profile of the positive particles $n(x)$ for $L=500$,
         $M=200$ and $q=0.45$ (Plot of the Eq.(12)). The negative particle is at $x=0$.}
\end{figure}
\section{Comparison and concluding remarks}
In this paper we investigated a recently introduced model [8,9] in the
special limit of infinite system size and infinite number of positively
charged particles in the presence of one negative particle. In this limit,
the model shows both phase transition and spatial condensation of particles
in one-dimension.
The phase transition takes place at $q_c=2\rho$, between a power-law and a
jammed phase. The speed of the positive and the negative particles are not
differentiable at the transition point $q_c=2\rho$. In the neutral case,
when the density of the positive and
negative particles are equal, the model shows a first order phase
transition at $q_c=\frac{1+6\rho}{1+2\rho}$, where the density of particles
as a function of fugacity has a finite jump [8,10]. Very recently it has been
shown that the transition point can be determined, with great
accuracy, using Yang-Lee theory in the non-equilibrium context [10].
In [10] the author presents a numerical study, using the grand canonical
ensemble, but in contrast to
the original Yang-Lee paper does not give a proof of the existence or
nonexistence of the phase transition.
Although it was claimed that neither in the frame-work of a grand
canonical nor canonical ensemble such transition exists [11].
Our exact calculations admit that in the limit $(L,M
\rightarrow \infty$ and $\frac{M}{L-1}=\rho$ being fixed) the phase transition
exists and in comparison to the neutral case [8]
the transition point is shifted. In the neutral case the current
of both positive and negative particles is zero for $q < 1$ where the
particle segregation exists. Here, as can be seen from Figures 1 and
2, the speed of particles never drops to zero for $q > 0$. In our model
the presence of a single negative particle produces a macroscopic shock
for $q \leq q_c$. In the power-law phase $q \geq q_c$ the density profile
of positive particles has an exponential behaviour with correlation length
$\xi$ which diverges at the transition point. This implies a second order
phase transition at this point. \\
The next step could be analytical study of this model with unequal but finite
densities of positive $\rho_{+}$ and negative particles $\rho_{-}$. An
approach can be considering the model with open boundaries where the
injection
and extraction rates control the bulk density.
Work in this direction is in progress [12].\\

{\bf{Appendix: Formulas for $Z_{L,M}$, $V_{+}$ and $V_{-}$}}\\

In the following we show that the normalization factor $Z_{L,M}$ can be
calculated directly using the algebra (3-5). First, we note that one can
split (6) in two terms as
\begin{equation}
Z_{L,M}=Tr(G_{L-1,M}EA)+Tr(G_{L-1,M-1}DA).
\end{equation}
Now using (4) and (5) it is not difficult to see that (15) can be
written as
\begin{equation}
Z_{L,M}=C_{L-2}^{M} Tr(E^{L-M-1}A)+Tr(G_{L-1,M-1}DA).
\end{equation}
This can be done by noting that the first term in (15) can be written as
$$
Tr(G_{L-1,M}EA)=Tr( \sum_{ \{\tau_i=0,1\}} \delta(M- \sum_{i=1}^{L-2} \tau_i)
\prod_{i=1}^{L-2}( \tau_i D+(1- \tau_i)E)EA)=
$$
$$
\sum_{ \{\tau_i=0,1\}} \delta(M- \sum_{i=1}^{L-2} \tau_i)
Tr(\prod_{i=1}^{L-2}( \tau_i D+(1- \tau_i)E)EA)=
$$
$$
\sum_{ \{\tau_i=0,1\}}\delta(M- \sum_{i=1}^{L-2} \tau_i)Tr(E^{L-M-2}EA)
$$
which gives the first term in (16).
One can expand the second term in (16) repeatedly as follows
$$
Tr(G_{L-1,M-1}DA)=Tr(G_{L-2,M-1}EDA)+Tr(G_{L-2,M-2}DDA)=
$$
$$
C_{L-3}^{M-1}Tr(E^{L-M-2}DA)+Tr(G_{L-2,M-2}D^2A)
$$
to obtain the following compact formula for the partition function
\begin{equation}
Z_{L,M}=\sum_{i=0}^{M}Tr(E^{L-M-1}D^iA)C_{L-i-2}^{M-i}.
\end{equation}
The expression (17) can easily be calculated using the algebra (3-5)
$$
f_i \equiv Tr(E^{L-M-1}D^iA)=
Tr(E^{L-M-1}D^{i-1}\frac{1}{q}(AD+A+D))=
 \frac{1}{q}(2f_{i-1}+Tr(E^{L-M-1})).
$$
Solving the above difference equation gives
\begin{equation}
f_i=\frac{(q-3)(\frac{2}{q})^i+1}{q-2}Tr(E^{L-M-1}) \ \  ,  \ \ i=0...M.
\end{equation}
Since $Z_{L,M}$ enters to both denominator and numerator of physical
quantities, the same common factor $Tr(E^{L-M-1})$ cancels from all
formulas so the results are independent of its value. For simplicity we
set $Tr(E^{L-M-1})=1$. \\
The current of the positive particles is defined as
\begin{equation}
J_{+}=P(\tau_i=1,\tau_{i+1}=0)+qP(\tau_i=1,\tau_{i+1}=2)-
P(\tau_{i+1}=1,\tau_{i}=2)
\end{equation}
in which $P(\tau_i=m,\tau_j=n)$ is the probability of finding the system in
configuration $\{ C \}$ provided that a particle of kind $m$ is
at site $i$ and a particle of kind $n$ at site $j$. Now we can write (19)
in term of the conditional probability (2) as follows
$$
J_{+}=P(\tau_i=1,\tau_{i+1}=0)+qP(\tau_i=1|\tau_{i+1}=2)P(\tau_{i+1}=2)-
P(\tau_{i+1}=1|\tau_{i}=2)P(\tau_{i}=2)=$$
$$
P(\tau_i=1,\tau_{i+1}=0)+
\frac{1}{L} (
q\frac{Tr(G_{L-1,M-1}DA)}{Tr(G_{L,M}A)}-
\frac{Tr(G_{L-1,M-1}AD)}{Tr(G_{L,M}A)})
$$
in which we have used $P(\tau_i=2)=\frac{1}{L}$. Using the algebra (3-5)
we obtain
$$
J_{+}=P(\tau_i=1,\tau_{i+1}=0)+
\frac{1}{L}\frac{Tr(G_{L-2,M-1}(A+D))}{Tr(G_{L,M}A)}=
$$
\begin{equation}
P(\tau_i=1,\tau_{i+1}=0)+\frac{1}{L}\frac{Z_{L-1,M-1}+C_{L-2}^{M-1}}{Z_{L,M}}
\end{equation}
in which we have $Tr(G_{L-2,M-1}A)=Z_{L-1,M-1}$ and
$$
Tr(G_{L-2,M-1}D)=
\sum_{ \{\tau_i=0,1\}}\delta(M-1- \sum_{i=1}^{L-2} \tau_i)
Tr(\prod_{i=1}^{L-2}( \tau_i D+(1- \tau_i)E)D)=
$$
$$
\sum_{ \{\tau_i=0,1\}}\delta(M-1- \sum_{i=1}^{L-2} \tau_i)
Tr(E^{L-M-2})=C_{L-2}^{M-1}
$$
The first term in (20) can be written as
$$P(\tau_i=1,\tau_{i+1}=0)=\sum_{k=1,k \neq
i,i+1}^{L}P(\tau_i=1,\tau_{i+1}=0 | \tau_k=2)P(\tau_k=2)=$$
$$ \frac{1}{L} \sum_{k=2}^{L-1} P( \tau_k=1, \tau_{k+1}=0 | \tau_1=2)=
\frac{1}{L}
\sum_{k=2}^{L-1} \sum_{p=0}^{M-1}\frac{Tr(G_{j,p}DEG_{L-j-1,M-p}A)}
{Tr(G_{L,M}A)}= $$
$$
\frac{(L-M-1)Z_{L-1,M-1}}{LZ_{L,M}}.
$$
Putting the last term in (20) gives
$$
J_{+}=\frac{(L-M)Z_{L-1,M-1}+C_{L-2}^{M-1}}{LZ_{L,M}}.
$$
Now we can calculate the speed of the positive particles
\begin{equation}
V_{+}=\frac{L}{M}J_{+}=\frac{(L-M)Z_{L-1,M-1}+C_{L-2}^{M-1}}{MZ_{L,M}}
\end{equation}
which is exactly the expression (8).
The speed of the negative particle is also defined as follows
$$
V_{-}=P(\tau_i=0|\tau_{i+1}=2)+qP(\tau_i=1|\tau_{i+1}=2)-
P(\tau_{i+1}=1|\tau_{i}=2).
$$
Now using the expression (2) for the conditional probability and the
algebra (3-5) we obtain
$$ V_{-}=\frac{Tr(G_{L-1,M}EA)}{Tr(G_{L,M}A)} +
q\frac{Tr(G_{L-1,M}DA)}{Tr(G_{L,M}A)}
-\frac{Tr(G_{L-1,M}AD)}{Tr(G_{L,M}A)}= $$
\begin{equation}
\frac{Tr(G_{L-1,M}EA)}{Tr(G_{L,M}A)}+\frac{Z_{L-1,M-1}+C_{L-2}^{M-1}}{Z_{L,M}}=
\frac{Z_{L-1,M-1}+C_{L-2}^{M}+C_{L-2}^{M-1}}{Z_{L,M}}.
\end{equation}
{ \large \bf Acknowledgement: } \\
I would like to thank V. Karimipour both for reading the manuscript and his comments.
I also acknowledge R. W. Sorfleet for his useful help during the preparation of this work.

\end{document}